\documentclass[11pt]{article}

\usepackage{amsfonts}
\usepackage{amscd}
\usepackage{epsfig}
\usepackage{amsmath}
\usepackage{amssymb}
\usepackage{latexsym}

\newcommand {\newsection}{\setcounter{equation}{0}\section}

\def\grafa#1{\begin{center} \epsfig{file=#1, width=7.7cm}
\quad}
\def\grafb#1{\epsfig{file=#1, width=7.7cm} \end{center}}

\def\bin#1#2{\pmatrix{ #1 \cr #2 \cr }}
\def\bin#1#2{\left(\begin{array}{c}#1\\#2\end{array}\right)}
\def\f{\frac}
\def\d{\partial}

\begin{document}

\title{A bi-Hamiltonian approach to the Sine-Gordon and 
Liouville hierarchies}
\author{Paolo Lorenzoni}
\maketitle
\begin{center}   
Department of Mathematics and Applications, University of Milano Bicocca
\\via degli Arcimboldi 8, 20126 Milano, Italy\\{\tt 
lorenzoni@matapp.unimib.it}
\end{center}
\vskip 0.3cm

\noindent 
{\bf Abstract}: In this paper we study the sine-Gordon and 
the Liouville hierarchies in 
laboratory coordinates from a bi-Hamiltonian point of view. Besides the 
well-known local structure these hierarchies possess a second 
compatible non-local Poisson structure.  

\vspace{0.3cm}



\newsection{Introduction} 

In \cite{zhang} Dubrovin and Zhang started a classification program of 
integrable evolutionary PDEs based on the bi-Hamiltonian geometry.

As a first step in the realization of this program they studied
 bi-Hamiltonian  hierarchies of PDEs of the form
\begin{equation}
u^{i}_{t}= A^{i}_{j}(u)u^{j}_{x}
+\sum_{k=1}^{\infty}\epsilon^{k}P_{k}(u,u_{x},...)
\label{hierarchy}
\end{equation}

where $P_{k}(u,u_{x},...)$ are differential polynomials.

A central role in their approach is played by the bi-Hamiltonian structure
 of the dispersionless limit ($\epsilon=0$) of the hierarchy
 (\ref{hierarchy}). Such structure consists of a pair of compatible $local$ 
Poisson brackets of hydrodynamic type (see \cite{zhang}).

Very important examples of integrable PDEs such as the KdV equation,
 the continuous limit of the Toda lattice and,
 after a suitable change of coordinates, the NLS
 equation belong to the class of hierarchies considered by Dubrovin and Zhang.

An open problem is to understand how to extend
 the classification scheme of \cite{zhang} onto more wide class of 
integrable systems
 not admitting dispersionless limit.
 The first point to clarify 
is which kind of bi-Hamiltonian structure is involved in this case.

In this paper we condider two important examples of this class of systems:
 the sine-Gordon equation, one of the first PDE solved by
 inverse scattering transform (see \cite{AKNS} and \cite{FT})
 and the Liouville equation:

\begin{eqnarray}
&&\omega_{tt}-\omega_{xx}=-\sin{\omega}\\
&&\phi_{tt}-\phi_{xx}=-e^{\phi}
\end{eqnarray}
 
(in laboratory coordinates).

Starting from the results of
a paper of Boiti,Leon and Pempinelli (\cite{boiti}), 
we show how to construct their bi-Hamiltonian structures.  

In sine-Gordon case 
such structure consists of the following compatible pair of Poisson 
bivectors

\begin{eqnarray}
&&P_{1}=
\begin{pmatrix}
\d&-2i\cr
2i&0\cr
\end{pmatrix}
\label{PoisSG1}
\end{eqnarray}

and 

\begin{eqnarray}
&&P_{2}=
\begin{pmatrix}
S^{1}_{(0)}\d^{-1}S^{1}_{(0)}+(\frac{u^{2}}{2}+s)\d+\d(\frac{u^{2}}{2}+s)
-\frac{\d^{3}}{4}&S^{1}_{(0)}\d^{-1}S^{2}_{(0)}-2i(u^{2}+s)+\frac{i\d^{2}}{2}\cr
S^{2}_{(0)}\d^{-1}S^{1}_{(0)}+2i(u^{2}+s)-\frac{i\d^{2}}{2}&S^{2}_{(0)}
\d^{-1}S^{2}_{(0)}-\d\cr
\end{pmatrix}
\label{PoisSG2}
\end{eqnarray}

with $S^{1}_{(0)}=i(\f{i}{8}\sin{\omega}+u_{x})$, $S^{2}_{(0)}=-2u$ and $s=s_{0}\cos{\omega}$, 
while in the Liouville case we obtain the pair

\begin{eqnarray}
&&P'_{1}=
\begin{pmatrix}
\d&2\cr
-2&0\cr
\end{pmatrix}
\label{PoisL1}
\end{eqnarray}

and $P'_{2}=$

\begin{eqnarray}
\f{1}{16}
\begin{pmatrix}
\d-S^{'1}_{(0)}\d^{-1}S^{'1}_{(0)}
+(8u^{2}+e^{\phi})\d+\d(8u^{2}+e^{\phi})
-4\d^{3}&
-S^{'1}_{(0)}\d^{-1}S^{'2}_{(0)}+32u^{2}+2e^{\phi}+2-8\d^{2}\cr
-S^{'2}_{(0)}\d^{-1}S^{'1}_{(0)}-32u^{2}-2e^{\phi}-2+8\d^{2}&
-S^{'2}_{(0)}\d^{-1}S^{'2}_{(0)}+16\d\cr
\end{pmatrix}
\label{PoisL2}
\end{eqnarray}

where $S^{'1}_{(0)}=\f{\exp{\phi}}{2}+4u_{x}$ and $S^{'2}_{(0)}=-8u$.

\vspace{.3 cm}

Following the Lenard-Magri scheme (\cite{magri}) one can construct 
recursively the integrals of motion. 
In the sine-Gordon case these integrals usually appear
 in two distinct families corresponding to 
the asymptotic expansion of the reduced monodromy matrix $T(\lambda)$ 
around $\lambda=0$ and $\lambda=\infty$ (\cite{FT2}) .In the 
bi-Hamiltonian framework  these two families are arranged in a single
 hierarchy, infinite in both directions.

\vspace{.3 cm}
Our observation is that, in both cases, the bivectors (\ref{PoisSG2})
 and (\ref{PoisL2}) are $weakly$ $nonlocal$
according to the definition of Maltsev and Novikov 
 (see \cite{maltsev}).

This class of nonlocal Poisson bivectors appeared only recently in
the mathematical literature  as 
a generalization of certain nonlocal Poisson brackets
 of hydrodynamic type (see \cite{ferapontov}, 
\cite{mokhov} and \cite{mokhov2}).

Nevertheless the list of $(1+1)$-integrable systems admitting such kind 
of Poisson structures contains, besides the sine-Gordon and Liouville
 equations, many other well-known integrable systems as KdV
 (considering higher Poisson structures) and NLS (see 
\cite{maltsev}).

To the best of our knowledge the bi-Hamiltonian structure of the
 Liouville hierarchy was unknown, while the bi-Hamiltonian structure of 
the sine-Gordon hierarchy appeared the first time in the paper \cite{fuch1}
 (see remark 5 in section 3).

\section{Preliminary definitions}

Let $L$ be the space of maps from the real line to some $n$-dimensional   
 manifold $M$ endowed with local coordinates $u^{1},...,u^{n}$ and 
let $\mathcal{A}$ be the space of differential polynomials in $u^{i,s}$,
that is
\begin{equation}
f\in \mathcal{A}\Leftrightarrow f=\sum
f_{i_{1},s_{1};...;i_{m},s_{m}}(u)u^{(i_{1},s_{1})}...u^{(i_{m},s_{m})},
\end{equation}

where
\begin{eqnarray*}
&&u^{(i,1)}=u^{i}_{x}\\
&&u^{(i,2)}=u^{i}_{xx}\\
&&...
\end{eqnarray*}

The role of the functions on $\mathcal{L}$ is played by the local
functionals
\begin{equation*}
I=\int_{-\infty}^{+\infty}f(u(x),u_{x},u_{xx},...)dx
\end{equation*}

where $f\in \mathcal{A}$.

\newtheorem{de}{Definition}
\begin{de}
A (non local) multivector $X$ is a formal infinite sum of the type
\begin{eqnarray*}
X=X^{i_{1},s_{1};,...;i_{k},s_{k}}(x_{1},...,x_{k};u(x_{1}),...,u(x_{k}),u_{x}
(x_{1}),...)
\frac{\partial}{\partial
u^{i_{1},s_{1}}(x_{1})}\wedge...\wedge
\frac{\partial}{\partial u^{i_{k},s_{k}}(x_{k})}
\end{eqnarray*}
\end{de}

where the coefficients satisfy the skew-symmetry condition with respect to
simultaneous permutations
\begin{eqnarray*}
i_{p},s_{p},x_{p}\leftrightarrow i_{q},s_{q},x_{q}
\end{eqnarray*}  

\begin{de}
A k-vector is called translation invariant if
\begin{eqnarray*}
X^{i_{1},s_{1};...;i_{k},s_{k}}(x_{1},...,x_{k};u(x_{1}),...,u(x_{k}),...)=
\d_{x_{1}}^{s_{1}}...\d_{x_{k}}^{s_{k}}
X^{i_{1},...,i_{k}}(x_{1},...,x_{k};u(x_{1}),...,u(x_{k}),...)
\end{eqnarray*}

where $X^{i_{1},...,i_{k}}(...)$ means $X^{i_{1},0;...;i_{k},0}(...)$ and
for any $t$
\begin{eqnarray*}
X^{i_{1},...,i_{k}}(x_{1}+t,...,x_{k}+t;u(x_{1}),...,u(x_{k}),...)=
X^{i_{1},...,i_{k}}(x_{1},...,x_{k};u(x_{1}),...,u(x_{k}),...)
\end{eqnarray*}
\end{de}

It follows from this definition that a translation invariant multi-vector
field is completely characterized by the ``components''

\begin{equation*}
X^{i_{1}...i_{k}}:=X(x_{1},...,x_{k};u(x_{1}),...,u(x_{k}),...)
\end{equation*}

\begin{de}
A bivector $P$ is called Poisson bivector if and only if 
\begin{equation}
[P,P]_{schouten}=0
\end{equation}
\end{de}

We recall that  the Schouten bracket of two translation 
invariant bivectors is given by the formula (see \cite{zhang}):
\begin{eqnarray}
&&\nonumber [P,Q]^{ijk}_{xyz}=\f{1}{2}\sum_{s}\left(
\f{\d P^{ij}_{xy}}{\d u^{l,s}(x)}\d_{x}^{s}Q^{lk}_{xz}
+\f{\d Q^{ij}_{xy}}{\d u^{l,s}(x)}\d_{x}^{s}P^{lk}_{xz}
+\f{\d P^{ij}_{xy}}{\d u^{l,s}(y)}\d_{y}^{s}Q^{lk}_{yz}
+\f{\d Q^{ij}_{xy}}{\d u^{l,s}(y)}\d_{y}^{s}P^{lk}_{yz}\right.\\
&&\nonumber \left.+\f{\d P^{ki}_{zx}}{\d u^{l,s}(z)}\d_{z}^{s}Q^{lj}_{zy}
+\f{\d Q^{ki}_{zx}}{\d u^{l,s}(z)}\d_{z}^{s}P^{lj}_{zy}
+\f{\d P^{ki}_{zx}}{\d u^{l,s}(x)}\d_{x}^{s}Q^{lj}_{xy}
+\f{\d Q^{ki}_{zx}}{\d u^{l,s}(x)}\d_{x}^{s}P^{lj}_{xy}
+\f{\d P^{jk}_{yz}}{\d u^{l,s}(y)}\d_{y}^{s}Q^{li}_{yx}+\right.\\
&&\left.+\f{\d Q^{jk}_{yz}}{\d u^{l,s}(y)}\d_{y}^{s}P^{li}_{yx}
+\f{\d P^{jk}_{yz}}{\d u^{l,s}(z)}\d_{z}^{s}Q^{li}_{zx}
+\f{\d Q^{jk}_{yz}}{\d u^{l,s}(z)}\d_{z}^{s}P^{li}_{zx}\right)
\label{formula}
\end{eqnarray}

A Poisson bivector defines a Poisson bracket. Indeed we have the following
\begin{de}
The Poisson bracket of two
local functionals $I_{1}$ ,$I_{2}$ is given by the formula
\begin{equation}
\{I_{1},I_{2}\}:=\int_{-\infty}^{+\infty}\int_{-\infty}^{+\infty}\f{\delta
I_{1}}{\delta
u^{i}(x)}P^{xy}\f{\delta I_{2}}{\delta u^{j}(y)}dxdy
\label{PB}
\end{equation} 
\end{de}

\begin{de}
A pair of bivetcors $(P_{1},P_{2})$ defines a bi-Hamiltonian structure if 
and only if 
\begin{equation}
[P_{1},P_{1}]_{Schouten}=[P_{1},P_{2}]_{Schouten}=[P_{2},P_{2}]_{Schouten}=0.
\end{equation}
\end{de}

\begin{de}
A Poisson bivector $P$ of components $P^{ij}_{xy}$ (and the corresponding 
Poisson bracket)
is called local if it can be written as:
\begin{equation}
P^{ij}_{xy}=\sum_{k}B^{ij}_{k}(u,u_{x},....)\delta^{(k)}(x-y)
\label{local}
\end{equation}
\end{de}  

By substituting (\ref{local}) in (\ref{PB}) we get
\begin{equation}
\{I_{1},I_{2}\}:=\int\f{\delta I_{1}}{\delta
u^{i}(x)}P^{ij}\f{\delta I_{2}}{\delta u^{j}(x)}dx
\end{equation}

where
\begin{equation}
P^{ij}=\sum_{k}B^{ij}_{k}(u,u_{x},....)\d^{k}
\end{equation}

($\d:=\f{d}{dx}$)

\begin{de}
A Poisson bivector $P$ of components $P^{ij}_{xy}$ (and the corresponding 
Poisson bracket)
is called weakly non-local (see \cite{maltsev}) if it can be written as:
\begin{equation}
P^{ij}_{xy}=\sum_{k}B^{ij}_{k}(u,u_{x},....)\delta^{(k)}(x-y)
+\sum_{k,l}e_{kl}S^{i}_{(k)}(u,u_{x},...)\nu(x-y)S^{j}_{(l)}(u,u_{y},...)
\label{nonlocal}
\end{equation}
where $\nu(x-y)$ is the step function ($=\f{1}{2}$ if $x-y$ is positive
and $=-\f{1}{2}$ if $x-y$ is negative).
The constants $e_{kl}=e_{lk}$ define a quadratic form in the space
generated by the flows $S^{i}_{(k)}(u,u_{x},...)$.
\end{de}

By substituting (\ref{nonlocal}) in (\ref{PB}) we get
\begin{equation}
\{I_{1},I_{2}\}:=\int_{-\infty}^{+\infty}\f{\delta I_{1}}{\delta
u^{i}(x)}P^{ij}\f{\delta I_{2}}{\delta u^{j}(x)}dx
\end{equation}

where
\begin{equation}
P^{ij}=\sum_{k}B^{ij}_{k}(u,u_{x},....)\d^{k}+\sum_{k,l}e_{kl}
S^{i}_{(k)}(u,u_{x},...)\d^{-1}S^{j}_{(l)}(u,u_{x},...)
\end{equation}  

and, by definition
\begin{equation}
\d^{-1}(.):=\f{1}{2}\left(\int_{-\infty}^{x}-\int_{x}^{+\infty}\right)dx(.)
\label{invdif}
\end{equation}   

Later on we will call local and weakly non local bivector also the
corresponding differential and integro-differential operator.

Similarly it is possible to define the class of weakly nonlocal symplectic 
structures (see \cite{maltsev}).

\begin{de}
Let $P$ be an invertible Poisson bivector. Then $P^{-1}$ is a weakly 
nonlocal symplectic structure if and only if it can be written in the 
following way:
\begin{equation}
(P^{-1})_{ij}=\sum_{k}B_{(k)ij}(u,u_{x},....)\d^{k}+\sum_{k,l}e_{kl}
Q^{i}_{(k)}(u,u_{x},...)\d^{-1}Q^{j}_{(l)}(u,u_{x},...)
\end{equation}
\end{de}

\section{The sine-Gordon hierarchy}
\subsection{Useful results}

$\bullet$ The starting point of the paper \cite{boiti} is the hierarchy of 
equations 
represented by the equation of zero curvature:
\begin{equation}
U_{t}-V_{x}+\left[U,V\right]=0
\label{zeroc}
\end{equation}

where (see \cite{boiti} for details)
\begin{eqnarray}
&&U=-i\lambda\sigma_{3}+u\sigma_{1}+(\f{i}{\lambda})(s\sigma_{3}+iv\sigma_{2})\\
&&V=\sum_{j=0}^{n}V_{j}\lambda^{n-j}+\sum_{j=0}^{p}W_{j}\lambda^{j-p-1}
\end{eqnarray}

$\sigma_{j}$ are the Pauli matrices and the functions $u(x)$, $v(x)$ and 
$s(x)$ have the following behavior for large $|x|$:
\begin{eqnarray*}
&&lim_{|x|\rightarrow\infty}u(x)=0\\
&&lim_{|x|\rightarrow\infty}v(x)=0\\ 
&&lim_{|x|\rightarrow\infty}s(x)=s_{0}
\end{eqnarray*}

The sine-Gordon hierarchy corresponds to the reduction of this hierarchy 
on the surface
\begin{equation}
s^{2}-v^{2}=s_{0}^{2}
\label{vincolo1}
\end{equation} 

\vspace{.3 cm}

$\bullet$ The equations of this hierarchy can be written as
\begin{eqnarray}
&&     
\begin{pmatrix}
u_{t}\cr
v_{t}\cr
\end{pmatrix}=-i\sum_{j=0}^{m}\mu_{j}^{(+)}J\f{\delta H_{2j+1}^{(+)}}{\delta q}+
i\sum_{j=0}^{k}\mu_{j}^{(-)}J\f{\delta H_{2j+1}^{(-)}}{\delta q}
\end{eqnarray} 

where
\begin{eqnarray}
&&
\f{\delta}{\delta q}=
\begin{pmatrix}
\f{\delta}{\delta u}\cr
\f{\delta}{\delta v}+\f{v}{s}\f{\delta}{\delta s}
\end{pmatrix}
\end{eqnarray} 

and

\begin{eqnarray}
&&
J=\begin{pmatrix}
\d&2s(x)\cr
-2s(x)&0\cr
\end{pmatrix}
\label{J}
\end{eqnarray} 

The first ``Hamiltonians'' are:
\begin{eqnarray}
&&H^{(+)}_{1}=i\int_{-\infty}^{+\infty}dx\left(s-s_{0}+\f{1}{2}u^{2}\right)\\
&&H^{(-)}_{1}=-i\int_{-\infty}^{+\infty}dx\left(s-s_{0}+\f{1}{2}u^{2}
+\f{v_{x}}{8s}\left(\f{v_{x}}{s}+4u\right)\right)
\end{eqnarray}

The sine-Gordon equation corresponds to the choice $m=k=0$, $\mu^{(+)}_{0}
=\mu^{(-)}_{0}=1$, $\mu^{(+)}_{j}=\mu^{(-)}_{j}=0$ for $j$ different from 
0. In fact by straightforward calculation one gets:
\begin{eqnarray}
&&
\begin{pmatrix}
u_{t}\cr
v_{t}\cr
\end{pmatrix}=
\begin{pmatrix}
u_{x}+4v\cr
-v_{x}-4us\cr
\end{pmatrix}
\label{SG}
\end{eqnarray}

On the surface (\ref{vincolo1}) $v$ and $s$ can be written as functions 
of $\omega$:
\begin{eqnarray}
&&v=is_{0}\sin{\omega}\\
\label{v}
&&s=s_{0}\cos{\omega}
\label{s}
\end{eqnarray}

By inserting (\ref{v}) and (\ref{s}) in the system (\ref{SG}) 
one obtains sine-Gordon equation
\begin{equation}
\omega_{tt}-\omega_{xx}=-16s_{0}\sin{\omega}
\end{equation}

\vspace{.3 cm} 

$\bullet$ The recursion operators for this hierarchy are given by the formulae
\begin{eqnarray}
N=\begin{pmatrix}
(2v+u_{x})\d^{-1}u+u^{2}+s-\frac{\d^{2}}{4}&(2v+u_{x})\d^{-1}\frac{v}{s}
+\frac{uv}{s}-\frac{\d}{2}\cr
-2su\d^{-1}u+\frac{1}{2}s\d&
-2su\d^{-1}\frac{v}{s}+s\cr 
\end{pmatrix}
\label{N}
\end{eqnarray}   

\begin{eqnarray}
N^{-1}=\f{1}{s_{0}^{2}}\begin{pmatrix}
-v\d^{-1}w+s&
-2v\d^{-1}\left(\frac{v}{s}+\frac{1}{4}w\d\frac{1}{s}\right)
+\frac{1}{2}s\d\frac{1}{s}\cr
\frac{1}{2}s\left(-\d+w\d^{-1}w\right)&
sw\d^{-1}\left(\frac{v}{s}+\frac{1}{4}w\d\frac{1}{s}\right)
-\frac{1}{4}s\d^{2}\frac{1}{s}+s\cr
\end{pmatrix}
\label{NInv}
\end{eqnarray}  

where $w=\f{v_{x}}{s}+2u$.

\vspace{.3 cm} 

$\bullet$ The hierarchy is local, that is all the equations are 
differential. 

\subsection{From the recursion operator to the bi-Hamiltonian structure}
 
In this section we show how to construct the sine-Gordon bi-Hamiltonian 
structure starting from the pair $(J,NJ)$, where $NJ=$

\begin{eqnarray}
\begin{pmatrix}
-(2v+u_{x})\d^{-1}(2v+u_{x})+(\f{1}{2}u^{2}+s)\d+\d(\f{1}{2}u^{2}+s)
-\frac{\d^{3}}{4}&2(2v+u_{x})\d^{-1}us+2s(u^{2}+s)-\frac{\d^{2}}{2}s\cr
2us\d^{-1}(2v+u_{x})-2s(u^{2}+s)+\frac{s\d^{2}}{2}&
-4su\d^{-1}us+s\d s\cr
\end{pmatrix}
\label{NJ}
\end{eqnarray}

We begin with the following
\newtheorem{re}{Remark} 
\begin{re}
The system     

\begin{eqnarray}
&&
\begin{pmatrix}
u_{t}\cr
v_{t}\cr
\end{pmatrix}=
\begin{pmatrix}
P_{11}& P_{12}\cr
P_{21}& P_{22}\cr
\end{pmatrix}
\begin{pmatrix}
\f{\delta H}{\delta u}\cr
\f{\delta H}{\delta v}+\f{v}{s}\f{\delta H}{\delta s}\cr
\end{pmatrix}
\end{eqnarray}

is equivalent, on the cylinder (\ref{vincolo1}), 
to the Hamiltonian system

\begin{eqnarray}
&&
\begin{pmatrix}
u_{t}\cr
v_{t}\cr
s_{t}\cr
\end{pmatrix}=
\begin{pmatrix}
P_{11}& P_{12}& P_{12}\f{v}{s}\cr
P_{21}& P_{22}& P_{22}\f{v}{s}\cr
\f{v}{s}P_{21}& \f{v}{s}P_{22}& \f{v}{s}P_{22}\f{v}{s}\cr 
\end{pmatrix}
\begin{pmatrix}
\f{\delta H}{\delta u}\cr
\f{\delta H}{\delta v}\cr
\f{\delta H}{\delta s}\cr
\end{pmatrix}
\end{eqnarray} 

In fact the third equation
\begin{equation}
s_{t}=\f{v}{s}P_{21}\f{\delta H}{\delta u}+ \f{v}{s}P_{22}\f{\delta 
H}{\delta v}+ \f{v}{s}P_{22}\f{v}{s}\f{\delta H}{\delta s}=\f{v}{s}v_{t}
\end{equation}

is compatible with the constraint (\ref{vincolo1}).
\end{re}

Then we have to substitute the $2\times2$ matrices $J$ and $NJ$ 
with the corresponding  $3\times3$ matrices $\tilde{J}=$ 
\begin{eqnarray}
&& 
\begin{pmatrix}
J_{11}& J_{12}& J_{12}\f{v}{s}\cr
J_{21}& J_{22}& J_{22}\f{v}{s}\cr
\f{v}{s}J_{21}& \f{v}{s}J_{22}&
\f{v}{s}J_{22}\f{v}{s}\cr
\end{pmatrix}
\label{tildeJ} 
\end{eqnarray}

and $\tilde{NJ}=$ 

\begin{eqnarray}
&&
\begin{pmatrix}
(NJ)_{11}& (NJ)_{12}& (NJ)_{12}\f{v}{s}\cr
(NJ)_{21}& (NJ)_{22}& (NJ)_{22}\f{v}{s}\cr
\f{v}{s}(NJ)_{21}& \f{v}{s}(NJ)_{22}&
\f{v}{s}(NJ)_{22}\f{v}{s}\cr
\end{pmatrix}
\label{tildeNJ}
\end{eqnarray} 

The idea is that the sine-Gordon bi-Hamiltonian structure is the 
restriction of the pair $(\tilde{J},\tilde{NJ})$ on the surface 
(\ref{vincolo1}). By replacing 
the coordinates $(u,v,s)$ by the coordinates 
$(u,\omega,s_{0})$ where 

\begin{eqnarray}
&&v=is_{0}\sin{\omega}\\
&&s=s_{0}\cos{\omega}
\end{eqnarray}    

and taking into account that,
 after a change of coordinates $u^{i'}=u^{i'}(u^{k})$ a 
bivector $P^{ij}$ transforms as
\begin{equation}
P^{ij'}=\f{\d u^{i'}}{\d u^{k}}P^{kl}\f{\d u^{j'}}{\d u^{l}}
\end{equation}

we obtain
\begin{eqnarray*}
&&
\begin{pmatrix}
1&0&0\cr
0&-i\f{s}{s_{0}^{2}}&i\f{v}{s_{0}^{2}}\cr
0&-\f{v}{s_{0}}&\f{s}{s_{0}}\cr 
\end{pmatrix}
\begin{pmatrix}
J_{11}& J_{12}& J_{12}\f{v}{s}\cr
J_{21}& J_{22}& J_{22}\f{v}{s}\cr
\f{v}{s}J_{21}& \f{v}{s}J_{22}& 
\f{v}{s}J_{22}\f{v}{s}\cr
\end{pmatrix}
\begin{pmatrix}
1&0&0\cr
0&-i\f{s}{s_{0}^{2}}&-\f{v}{s_{0}}\cr
0&i\f{v}{s_{0}^{2}}&\f{s}{s_{0}}\cr
\end{pmatrix}=
\begin{pmatrix}
\d&-2i&0\cr
2i&0&0\cr
0&0&0\cr
\end{pmatrix}  
\end{eqnarray*}

and

\begin{eqnarray*}
&&
\begin{pmatrix}
1&0&0\cr
0&-i\f{s}{s_{0}^{2}}&i\f{v}{s_{0}^{2}}\cr
0&-\f{v}{s_{0}}&\f{s}{s_{0}}\cr
\end{pmatrix}
\begin{pmatrix}
(NJ)_{11}& (NJ)_{12}& (NJ)_{12}\f{v}{s}\cr
(NJ)_{21}& (NJ)_{22}& (NJ)_{22}\f{v}{s}\cr
\f{v}{s}(NJ)_{21}& \f{v}{s}(NJ)_{22}&
\f{v}{s}(NJ)_{22}\f{v}{s}\cr
\end{pmatrix}
\begin{pmatrix}
1&0&0\cr
0&-i\f{s}{s_{0}^{2}}&-\f{v}{s_{0}}\cr
0&i\f{v}{s_{0}^{2}}&\f{s}{s_{0}}\cr
\end{pmatrix}=
\end{eqnarray*} 
\begin{eqnarray*}
\begin{pmatrix}
-(2v+u_{x})\d^{-1}(2v+u_{x})+(\frac{u^{2}}{2}+s)\d+\d(\frac{u^{2}}{2}+s)
-\frac{\d^{3}}{4}&-2i(2v+u_{x})\d^{-1}u-2i(u^{2}+s)+\frac{i\d^{2}}{2}&0\cr
-2iu\d^{-1}(2v+u_{x})+2i(u^{2}+s)-\frac{i\d^{2}}{2}&4u\d^{-1}u-\d&0\cr
0&0&0\cr 
\end{pmatrix}
\end{eqnarray*}

>From these formulae it follows immediately that the restriction on the 
surface (\ref{vincolo1}) is given by the expressions
 (\ref{PoisSG1}) and (\ref{PoisSG2}).

Then we have  the following theorem.
\newtheorem{te}{Theorem}
\begin{te}
The pair $(\tilde{J},\tilde{NJ})$ and its restriction on the cylinder 
(\ref{vincolo1}) are bi-Hamiltonian structures.
\label{th9}
\end{te}

We omit the proof of this theorem consisting of a very long
 but straightforward calculation based on the formula (\ref{formula}).

The most difficult part of this kind of computation is to check
 whether certain expressions containing products of distributions vanish.
The trick is to use some identities between distributions like

\begin{eqnarray*}
&&f(y)\delta^{(s)}(x-y)=\sum_{q=0}^{s}\bin{s}{q}
f^{(q)}(x)\delta^{(s-q)}(x-y)\\
&&\delta(x-y)\delta(x-z)=\delta(y-x)\delta(y-z)=\delta(z-x)\delta(z-y)\\
&&\nu(y-z)\delta(z-x)=\d_{y}^{-1}\left(\delta(z-x)\delta(z-y)\right)
=\d_{y}^{-1}\left(\delta(x-y)\delta(x-z)\right)=-\nu(x-y)\delta(x-z)\\
&&\nu(x-y)\delta(y-z)=\d_{x}^{-1}\left(\delta(y-x)\delta(y-z)\right)
=\d_{x}^{-1}\left(\delta(z-x)\delta(z-y)\right)=-\nu(z-x)\delta(z-y)\\
&&\nu(z-x)\delta(x-y)=\d_{z}^{-1}\left(\delta(x-y)\delta(x-z)\right)   
=\d_{z}^{-1}\left(\delta(y-x)\delta(y-z)\right)=-\delta(y-x)\nu(y-z)
\end{eqnarray*}

\vspace{.3 cm}

>From the theorem it follows directly 

\newtheorem{co}{Corollary}
\begin{co} 
(\ref{PoisSG2}) is a weakly non-local Poisson  bivector.
\end{co}

\begin{re}
The two hierarchies $\{I_{n+1}\}:=\{-iH^{+}_{2n+1}\}$ and $\{I_{-n}\}:=
\{\f{i}{s_{0}}H^{-}_{2n+1}\}$ can be ``glued'' together. In fact we have:
\begin{equation}
P_{2}\delta I_{0}=\f{1}{s_{0}}iP_{2}\delta
H^{+}_{1}=-iP_{1}\delta H^{-}_{1}=P_{1}\delta I_{1}
\end{equation}

In other words we have the Lenard-Magri chain:
\begin{displaymath}
\begin{array}{rcl}
&&.\,.\,.\\
&\stackrel{P_{1}}{\nearrow}&\\
I_{-2}&&\\
& \stackrel{P_{2}}{\searrow} &\\
&& P_{2}\delta I_{-2}=P_{1}\delta I_{-1} \\
&\stackrel{P_{1}}{\nearrow}&\\
I_{-1}&&\\
& \stackrel{P_{2}}{\searrow} &\\
&& P_{2}\delta I_{-1}=P_{1}\delta I_{0} \\
&\stackrel{P_{1}}{\nearrow}&\\
I_{0}&&\\
& \stackrel{P_{2}}{\searrow} &\\
&& P_{2}\delta I_{0}=P_{1}\delta I_{1} \\
&\stackrel{P_{1}}{\nearrow} & \\
I_{1}&& \\
&\stackrel{P_{2}}{\searrow} & \\
&& P_{2}\delta I_{1}=P_{1}\delta I_{2} \\
&\stackrel{P_{1}}{\nearrow} & \\
I_{2}&& \\
&\stackrel{P_{2}}{\searrow} & \\
&&  .\,.\,.
\end{array}
\end{displaymath}
\end{re}

An immediate consequence of the last remark is the following
\begin{te}
The sine-Gordon equation belongs to the hierarchy
$\{H_{k}=s_{0}I_{k}+I_{k+1}\}_{k\in Z}$.

where
\begin{eqnarray}
&&H_{1SG}=\int\left(u^{2}+\frac{i}{2}u\omega_{x}-\frac{\omega_{x}^{2}}{8}
+2s_{0}\cos{\omega}\right)dx\\
\nonumber&&H_{2SG}=\\
\nonumber&&\int\f{1}{s_{0}}\left(\frac{1}{2}u^{2}+\frac{1}{2}iu\omega_{x}
-\frac{\omega_{x}^{2}}{8}+s_{0}\cos{\omega}\right)+\f{1}{s_{0}^{2}}
\left(\frac{s_{0}^{2}}{4}cos(2\omega)
-\frac{3s_{0}}{8}\cos{\omega}\omega_{x}^{2}-\frac{\omega_{xx}^{2}}{32}
+\frac{\omega_{x}^{4}}{128}\right.\\
&&\left.+\frac{s_{0}}{2}u^{2}\cos{\omega}+ius_{0}\cos{\omega}\omega_{x}
-\frac{i}{8}u\omega_{xxx}+\frac{1}{8}u_{x}^{2}-\frac{i}{16}u\omega_{x}^{3}
-\frac{3}{16}u^{2}\omega_{x}^{2}+\frac{i}{4}u^{3}\omega_{x}
+\frac{1}{8}u^{4}\right)dx
\label{hamiltonian}
\end{eqnarray}

The momentum integrals are given by the following expressions:

\begin{eqnarray}
&&P_{1SG}=-\f{1}{8}\int\left(4iu\omega_{x}-\omega_{x}^{2}\right)dx 
\\
\nonumber&&P_{2SG}=\\
\nonumber&&\int\f{1}{s_{0}}\left(\frac{1}{2}u^{2}+\frac{1}{2}iu\omega_{x}
-\frac{\omega_{x}^{2}}{8}+s_{0}\cos{\omega}\right)-\f{1}{s_{0}^{2}}
\left(\frac{s_{0}^{2}}{4}cos(2\omega)
-\frac{3s_{0}}{8}\cos{\omega}\omega_{x}^{2}-\frac{\omega_{xx}^{2}}{32}
+\frac{\omega_{x}^{4}}{128}\right.\\
&&\left.+\frac{s_{0}}{2}u^{2}\cos{\omega}+ius_{0}\cos{\omega}\omega_{x}
-\frac{i}{8}u\omega_{xxx}+\frac{1}{8}u_{x}^{2}-\frac{i}{16}u\omega_{x}^{3}
-\frac{3}{16}u^{2}\omega_{x}^{2}+\frac{i}{4}u^{3}\omega_{x}
+\frac{1}{8}u^{4}\right)dx
\label{momentum}
\end{eqnarray}

Moreover the Poisson bivectors $P_{1}$ and $P_{2}$ are invertible. More 
precisely:
\begin{eqnarray}                                                           
&&
P_{1}^{-1}=\begin{pmatrix}                                                             
0&-\f{i}{2}\cr
\f{i}{2}&-\f{\d}{4}\cr 
\end{pmatrix}
\label{P1inv}
\end{eqnarray}   

and $P_{2}^{-1}=\f{1}{s_{0}^{2}}\times$

\begin{eqnarray}
&&
\begin{pmatrix}
T_{1}\d^{-1}T_{1}+\f{1}{4}\d&T_{1}\d^{-1}T_{2}
-i\left(\f{s_{0}}{2}\cos{\omega}-\f{\d^{2}}{8}-\f{T_{1}^{2}}{2}\right)\cr
T_{2}\d^{-1}T_{1}
+i\left(\f{s_{0}}{2}\cos{\omega}-\f{\d^{2}}{8}-\f{T_{1}^{2}}{2}\right)
&T_{2}\d^{-1}T_{2}
+\left(\f{T_{1}^{2}}{4}-\f{s_{0}}{2}\cos{\omega}\right)\d
+\f{\left(\f{T_{1}^{2}}{4}-\f{s_{0}}{2}\cos{\omega}\right)_{x}}{2}\cr
\end{pmatrix}
\label{P2inv}
\end{eqnarray} 

with
\begin{eqnarray*}
&&T_{1}=\f{i}{2}\left(i\omega_{x}+2u\right)\\
&&T_{2}=-\left(is_{0}\sin{\omega}-\f{i}{4}\omega_{xx}-\f{1}{2}u_{x}\right)
\end{eqnarray*} 
\end{te}   

{\bf Proof}
By adding the Lenard-Magri chain for the Hamiltonians
$\{I_{k}\}_{k\in Z}$ multiplied by $s_{0}$ to the same chain shifted by
one we obtain
\begin{displaymath}
\begin{array}{rcl}
&&.\,.\,.\\
&\stackrel{P_{1}}{\nearrow}&\\
H_{-2}&&\\
& \stackrel{P_{2}}{\searrow} &\\
&& P_{2}\delta H_{-2}=P_{1}\delta H_{-1} \\
&\stackrel{P_{1}}{\nearrow}&\\
H_{-1}&&\\
& \stackrel{P_{2}}{\searrow} &\\
&& P_{2}\delta H_{-1}=P_{1}\delta H_{0} \\
&\stackrel{P_{1}}{\nearrow}&\\
H_{0}&&\\
& \stackrel{P_{2}}{\searrow} &\\
&& P_{2}\delta H_{0}=P_{1}\delta H_{1} \\
&\stackrel{P_{1}}{\nearrow} & \\
H_{1}&& \\
&\stackrel{P_{2}}{\searrow} & \\
&& P_{2}\delta H_{1}=P_{1}\delta H_{2} \\
&\stackrel{P_{1}}{\nearrow} & \\
H_{2}&& \\
&\stackrel{P_{2}}{\searrow} & \\
&&  .\,.\,.
\end{array}
\end{displaymath}      

In order to find the recursion operators of this hierarchy we need the 
following

\begin{re} 
Let $N$ and $M=N^{-1}$ be the $2\times 2$ recursion operator introduced 
above. If $M$ and $N$ satisfy the equations
\begin{eqnarray}
&&
\begin{pmatrix}
M_{11}&M_{12}\cr
M_{21}&M_{22}\cr
\end{pmatrix}
\begin{pmatrix}
J^{2}_{11}&J^{2}_{12}\cr
J^{2}_{21}&J^{2}_{22}\cr 
\end{pmatrix}=
\begin{pmatrix}
J^{1}_{11}&J^{1}_{12}\cr
J^{1}_{21}&J^{1}_{22}\cr
\end{pmatrix} 
\end{eqnarray} 

and 
   
\begin{eqnarray}
&&
\begin{pmatrix}
N_{11}&N_{12}\cr
N_{21}&N_{22}\cr
\end{pmatrix}
\begin{pmatrix}
J^{1}_{11}&J^{1}_{12}\cr
J^{1}_{21}&J^{1}_{22}\cr
\end{pmatrix}=
\begin{pmatrix}
J^{2}_{11}&J^{2}_{12}\cr
J^{2}_{21}&J^{2}_{22}\cr
\end{pmatrix}
\end{eqnarray}                    

then the $3\times 3$ matrices
\begin{eqnarray}
&&\tilde{M}=
\begin{pmatrix}
M_{11}&M_{12}&0\cr
M_{21}&M_{22}&0\cr
\f{v}{s}M_{21}&0&\f{v}{s}M_{22}\f{s}{v}
\end{pmatrix}  
\end{eqnarray}

and
\begin{eqnarray}
&&\tilde{N}=
\begin{pmatrix}
N_{11}&N_{12}&0\cr
N_{21}&N_{22}&0\cr
\f{v}{s}N_{21}&0&\f{v}{s}N_{22}\f{s}{v}
\end{pmatrix}
\end{eqnarray}

satisfy the equations:
\begin{equation}
\tilde{M}\tilde{J^{2}}=\tilde{J^{1}}
\label{mj}
\end{equation}

and

\begin{equation}
\tilde{N}\tilde{J^{1}}=\tilde{J^{2}}
\end{equation}

where 

\begin{eqnarray}
&&\tilde{J^{1}}= 
\begin{pmatrix}
J^{1}_{11}& J^{1}_{12}& J^{1}_{12}\f{v}{s}\cr
J^{1}_{21}& J^{1}_{22}& J^{1}_{22}\f{v}{s}\cr
\f{v}{s}J^{1}_{21}& \f{v}{s}J^{1}_{22}&
\f{v}{s}J^{1}_{22}\f{v}{s}\cr
\end{pmatrix}
\end{eqnarray} 

and

\begin{eqnarray}
&&\tilde{J^{2}}=
\begin{pmatrix}
J^{2}_{11}& J^{2}_{12}& J^{2}_{12}\f{v}{s}\cr
J^{2}_{21}& J^{2}_{22}& J^{2}_{22}\f{v}{s}\cr
\f{v}{s}J^{2}_{21}& \f{v}{s}J^{2}_{22}&
\f{v}{s}J^{2}_{22}\f{v}{s}\cr
\end{pmatrix}
\end{eqnarray}  

(in our case $\tilde{J^{1}}=(\ref{tildeJ})$ and 
$\tilde{J^{2}}=(\ref{tildeNJ})$)

\end{re}
Since $\tilde{J^{1}}$ and $\tilde{J^{2}}$ are not invertible in the
space $(u,v,s)$
we can try to express $\tilde{M}$ and $\tilde{N}$ in terms of
them only after
the restriction on the surface ($\ref{vincolo1}$).

Taking into account that after a change
of coordinates $u^{i'}=u^{i'}(u^{k})$ $\tilde{M}$ transforms
according to the rule
\begin{equation}
\tilde{M}^{i}_{j'}=\f{\d u^{i'}}{\d u^{k}}
\tilde{M}^{k}_{l}\f{\d u^{l}}{\d u^{j'}}
\end{equation}

it is easy to see that in the coordinates
$(u,\omega,s_{0})$ $\tilde{M}$ becomes:  
\begin{eqnarray*}
&&
\f{1}{s_{0}^{2}}
\begin{pmatrix}
s-v\d^{-1}w&\f{i}{2}s\d-2iv\d^{-1}\left(v+\f{1}{4}w\d\right)&*\cr
-\f{i}{2}\left(-\d+w\d^{-1}w\right)&s-\f{1}{4}\d^{2}+w\d^{-1}
\left(v+\f{1}{4}w\d\right)&*\cr
0&0&*\cr
\end{pmatrix}
\end{eqnarray*} 

The equations (\ref{mj}) in these coordinates becomes
\begin{eqnarray*}
&&
\f{1}{s_{0}^{2}}
\begin{pmatrix}
s-v\d^{-1}w&\f{i}{2}s\d-2iv\d^{-1}\left(v+\f{1}{4}w\d\right)&*\cr
-\f{i}{2}\left(-\d+w\d^{-1}w\right)&s-\f{1}{4}\d^{2}+w\d^{-1}
\left(v+\f{1}{4}w\d\right)&*\cr
0&0&*\cr
\end{pmatrix}
\begin{pmatrix} 
P_{2}^{11}&P_{2}^{12}&0\cr
P_{2}^{21}&P_{2}^{22}&0\cr
0&0&0\cr
\end{pmatrix}=
\begin{pmatrix}
P_{1}^{11}&P_{1}^{12}&0\cr
P_{1}^{21}&P_{1}^{22}&0\cr
0&0&0\cr
\end{pmatrix}    
\end{eqnarray*}

>From these equations it follows immediately that 
\begin{eqnarray*}
&&
M'=\f{1}{s_{0}^{2}}
\begin{pmatrix}
s-v\d^{-1}w&\f{i}{2}s\d-2iv\d^{-1}\left(v+\f{1}{4}w\d\right)\cr
-\f{i}{2}\left(-\d+w\d^{-1}w\right)&s-\f{1}{4}\d^{2}+w\d^{-1}
\left(v+\f{1}{4}w\d\right)\cr
\end{pmatrix}
\end{eqnarray*} 

satisfies the equation
\begin{equation}
M'P_{2}=P_{1}
\end{equation}

and therefore it is a recursion operator for the sine-Gordon hierarchy.

It is easy to check that the bivector (\ref{P1inv}) is the inverse of 
$P_{1}$ and that the bivector
($\ref{P2inv})=P^{-1}_{1}M'$ is the inverse of $P_{2}$.  

Finally from
\begin{eqnarray}
&&P_{1}\delta P_{1SG}=P_{2}\delta P_{2SG}\\
&&P_{1}\delta H_{1SG}=P_{2}\delta H_{2SG}
\end{eqnarray}

it follows immediately
\begin{eqnarray}
&&\delta P_{2SG}=P_{2}^{-1}P_{1}\delta P_{1SG}\\
&&\delta H_{2SG}=P_{2}^{-1}P_{1}\delta H_{1SG}
\end{eqnarray}

By straightforward calculation we get
\begin{eqnarray*}
&&\f{\delta H_{2SG}}{\delta u}=
\f{-1}{4}u_{xx}+is_{0}\cos{\omega}\omega_{x}-\f{3}{8}u\omega_{x}^{2}
-\f{3}{4}iu^{2}\omega_{x}+\f{1}{2}u^{3}+\f{1}{2}is_{0}\omega_{x}
+us_{0}\cos{\omega}+us_{0}-\f{1}{8}i\omega_{xxx}+\\
&&-\f{1}{16}\omega_{x}^{3}\\
&&\f{\delta H_{2SG}}{\delta \omega}= 
-is_{0}\cos{\omega}u_{x}-\f{1}{2}s_{0}^{2}\sin{2\omega}+\f{3}{8}iu\omega_{x}
\omega_{xx}+\f{3}{4}uu_{x}\omega_{x}+\f{3}{8}u^{2}\omega_{xx}
-s_{0}^{2}\sin{\omega}+\\
&&+\f{3}{4}s_{0}\cos{\omega}\omega_{xx}
+\f{1}{4}s_{0}\omega_{xx}-\f{1}{2}is_{0}u_{x}+\f{1}{8}iu_{xxx}
-\f{3}{8}s_{0}\sin{\omega}\omega_{x}^{2}+\f{3}{16}iu_{x}\omega_{x}^{2}
-\f{3}{32}\omega_{x}^{2}\omega_{xx}+\f{3}{4}iu^{2}u_{x}+\\
&&-\f{1}{16}\omega_{xxx}-\f{1}{2}u^{2}s_{0}\sin{\omega} 
\end{eqnarray*}

that is the differential of the Hamiltonian (\ref{hamiltonian}). 
Similarly we get (\ref{momentum}).

\begin{co}
The operator (\ref{P2inv}) defines a weakly nonlocal symplectic structure.
\end{co} 

\begin{re}
In order to obtain the usual symplectic structure of the sine-Gordon 
hierarchy

\begin{eqnarray}
&&\Theta_{1}=
\begin{pmatrix}
 0&1\cr
-1&0\cr
\end{pmatrix}
\label{symplectic}
\end{eqnarray}

it is sufficient  to make the change of variables $q=\omega$,
 $p=4iu-q_{x}$: in the new coordinates the Poisson bivector $(\ref{PoisSG1})$ 
becomes the usual one while the second one coincide with the Poisson bivector 
found by Fuchssteiner and Oevel (\cite{fuch1}).
\end{re}

\begin{re}
The geometrical structures (recursion operator, bi-Hamiltonian structure) of 
the sine-Gordon equation are well known in light-cone coordinates
 $(\xi=\f{t+x}{2},\eta=\f{t-x}{2})$ (see for example \cite{dorfman}).

Starting from these structures it is not obvious how to get their counterpart  
in laboratory coordinates.

A possible way was suggested by Fuchssteiner and Oevel in (\cite{fuch1}) and 
it is based on the properties of a particular class of time-dependent vector
 fields called master symmetries. In the case of integrable PDEs admitting a
 hamiltonian formulation with respect to some Poisson bivector $P$ a master 
symmetry  $\tau$ has the property that the Lie derivative $Lie_{\tau}P$ defines 
a new Poisson bivector compatible with $P$.

The main idea of the paper \cite{fuch1} is that the master symmetry of the
 sine-Gordon equation in light-cone coordinates can be modified in such 
a way that the new vector field coincides with it on the submanifold of
 the solutions and gives rise to an evolution symmetry with respect to
 the laboratory coordinates.

Such evolution symmetry is the master symmetry of the sine-Gordon hierarchy in 
laboratory coordinates. Therefore the Lie derivative of the symplectic structure
 (\ref{symplectic}) along it defines a second compatible Poisson bivector
 $\Theta_{2}$ and the 
operator $\Phi=\Theta_{2}\Theta^{-1}_{1}$ is an hereditary symmetry,
 in other words for any vector fields $A$ and $B$ it satisfies the equation:
\begin{equation}
\Phi^{2}[A,B]+[\Phi A,\Phi B] = \Phi \left([\Phi A,B]+[A,\Phi B]\right)
\end{equation}

The proof of the hereditaryness of the operator $\Phi$ is based on some computer
 algebra methods introduced in the paper \cite{fuch2}.
\end{re}

\section{The Liouville hierarchy}
\subsection{Useful results of \cite{boiti}}

$\bullet$ The Liouville hierarchy corresponds to the reduction
of the hierarchy of equations (\ref{zeroc}) on the plane in the space $(u,v,s)$
 given by the equation $s=v$.

\noindent
$\bullet$ The difference with respect to the sine-Gordon case is that
the recursion operator (\ref{N}) is no longer invertible and as a consequence
only the local Hamiltonians $\{H^{+}_{j}\}_{j\in N}$ and $H^{-}_{1}$
survive.

\subsection{The bi-Hamiltonian structure of the Liouville hierarchy} 

We have seen that the bi-Hamiltonian structure 
$(\ref{PoisSG1},\ref{PoisSG2})$ of the sine-Gordon hierarchy is simply the 
restriction of the pair $(\tilde{J},\tilde{NJ})$ on the surface 
$s^{2}-v^{2}=s_{0}^{2}$.
Similarly it is possible to obtain the bi-Hamiltonian structure of the Liouville 
hierarchy as restriction of the pair $(\tilde{J},\f{1}{16}\tilde{J}+\tilde{NJ})$ on the 
plane $s=v=\f{e^{\phi}}{16}$. The only difference in the Liouville case is that,
 in order to have a second $local$ hamiltonian, we have to choose
 $\f{1}{16}\tilde{J}+\tilde{NJ}$ instead of $\tilde{NJ}$.

\vspace{.3 cm}

{\bf Acknowledgments} \hspace{.3 cm} I would like to acknowledge 
Boris Dubrovin for useful discussions and for the reference \cite{boiti},
the starting point of the present work. I thank also Maciej Blaszak for the 
reference \cite{fuch1}.

\end{document}